\documentclass{elsart}

\usepackage{epsfig}

\usepackage{amsmath}

\newcommand{\be}{\begin{equation}}
\newcommand{\ee}{\end{equation}}
\newcommand{\bea}{\begin{eqnarray}}
\newcommand{\eea}{\end{eqnarray}}

 \def\bean{\begin{eqnarray*}}
 \def\eean{\end{eqnarray*}}

 \def\gsim{\mathrel{\rlap{\lower0.2em\hbox{$\sim$}}\raise0.2em\hbox{$>$}}}
 \def\ksim{\mathrel{\rlap{\lower0.2em\hbox{$\sim$}}\raise0.2em\hbox{$<$}}}

\begin{document}

\begin{frontmatter}
\title{Signatures of the strongly interacting QGP
in relativistic heavy-ion collisions}
\author[unif]{E.L.~Bratkovskaya},
\author[unif]{O.~Linnyk}, 
\author[unig]{V.P.~Konchakovski},
\author[unik]{M.I.~Gorenstein},
\author[unig]{W.~Cassing}
\address[unif]{Institut f\"ur Theoretische Physik, %
 Goethe-Universit\"{a}t Frankfurt am Main, 
  Germany}
\address[unig]{Institut f\"ur Theoretische Physik, %
 Justus-Liebig-Universit\"at Giessen, %
  Germany}
\address[unik]{
Bogolyubov Institute for Theoretical Physics, 
Ukraine}

\begin{abstract}
The transition from hadronic to partonic degrees of freedom in the
course of a relativistic heavy-ion collision is described by the
microscopic covariant Parton-Hadron-String Dynamics (PHSD) transport
approach. Studying Pb+Pb reactions from 40 to 158 A$\cdot$GeV and
comparing the PHSD results to those of the Hadron-String Dynamics
(HSD) approach without a phase transition to the QGP, we observe
that the existence of the partonic phase has a sizable influence on
the transverse mass distribution of final kaons due to the repulsive
partonic mean fields. Furthermore, we find a significant effect of
the QGP on the production of multi-strange antibaryons due to a
slightly enhanced $s\bar s$ pair production in the partonic phase
from massive time-like gluon decay and to a more abundant formation
of strange antibaryons in the hadronization process. Another
evidence for pre-hadronization dynamics is gained from a study of
di-jet correlations in Au+Au collisions at the top RHIC energy of
$\sqrt{s}=200$ GeV. Within the HSD transport approach, the reaction
of the hadronic medium to the jet energy loss is calculated.  In
comparison with the data of the STAR, PHOBOS and PHENIX
Collaborations differentially in azimuthal angle and pseudorapidity,
the HSD results do not show enough suppression for the `away-side'
jet. In addition, the HSD results exhibit neither a `Mach-cone'
structure for the angular distribution in the away-side jet nor the
`ridge' long-range rapidity correlations for the near-side jet as
observed by the STAR and PHOBOS Collaborations, thus suggesting a
partonic origin of these structures.
\end{abstract}

\end{frontmatter}

\vspace{-10pt}

The nature of confinement and the phase transition from a partonic
system of quarks, antiquarks and gluons -- a quark-gluon plasma
(QGP) -- to interacting hadrons, as occurring in relativistic
nucleus-nucleus collisions, is a central topic of modern high-energy
physics. In the present work, the dynamical evolution of the heavy
ion-collision is described by the PHSD transport
approach~\cite{CasBrat} incorporating the off-shell propagation of
the partonic quasi-particles according to~\cite{Juchem} as well as
the transition to resonant hadronic states (or strings). In
Section~2, we employ the PHSD approach -- described in Section~1 and
in more detail in~\cite{CasBrat,Cassing:2009vt} -- to strangeness
production in nucleus-nucleus collisions at moderate relativistic
energies, i.e. at SPS energies up to 160~A$\cdot$GeV. On the other
hand, one should look within the same approach at the whole spectrum
of observables with a special sensitivity to the partonic phase,
e.g. strangeness~\cite{Cassing:2009vt},
dileptons~\cite{Linnyk:2009nx}, charm~\cite{Linnyk:2008hp},
jets~\cite{Gallmeister:2004iz}, etc. In this spirit, we dedicate
Section~3 to a discussion of the medium reaction to the jet energy
loss in Au+Au collisions at the top RHIC energy of
$\sqrt{s}=200$~GeV.

\vspace{-20pt}
\section{The PHSD approach} \vspace{-10pt} \vspace{-10pt}

A consistent dynamical approach -- valid also for strongly
interacting systems -- can be formulated on the basis of the
Kadanoff-Baym equations \cite{KBaym,Sascha1} or off-shell transport
equations in phase-space representation, respectively
\cite{Juchem,Sascha1}. In the Kadanoff-Baym theory the field quanta
are described in terms of propagators with complex selfenergies.
Whereas the real part of the selfenergies can be related to
mean-field potentials, the imaginary parts  provide information
about the lifetime and/or reaction rates of time-like 'particles'
\cite{Andre}. Once the proper (complex) selfenergies of the degrees
of freedom are known, the time evolution of the system is fully
governed  by off-shell transport equations (as described in Refs.
\cite{Juchem,Sascha1}).

The PHSD approach is a microscopic covariant transport model that
incorporates effective partonic as well as hadronic degrees of
freedom and involves a dynamical description of the hadroni\-zation
process from partonic to hadronic matter \cite{CasBrat}. Whereas the
hadronic part is essentially equivalent to the conventional HSD
approach \cite{HSD} the partonic dynamics is based on the Dynamical
QuasiParticle Model (DQPM) \cite{Cassing06} which describes QCD
properties in terms of single-particle Green's functions (in the
sense of a two-particle irreducible approach) and leads to effective
strongly interacting partonic quasiparticles with broad spectral
functions as degrees of freedom.

\begin{figure}
{\psfig{figure=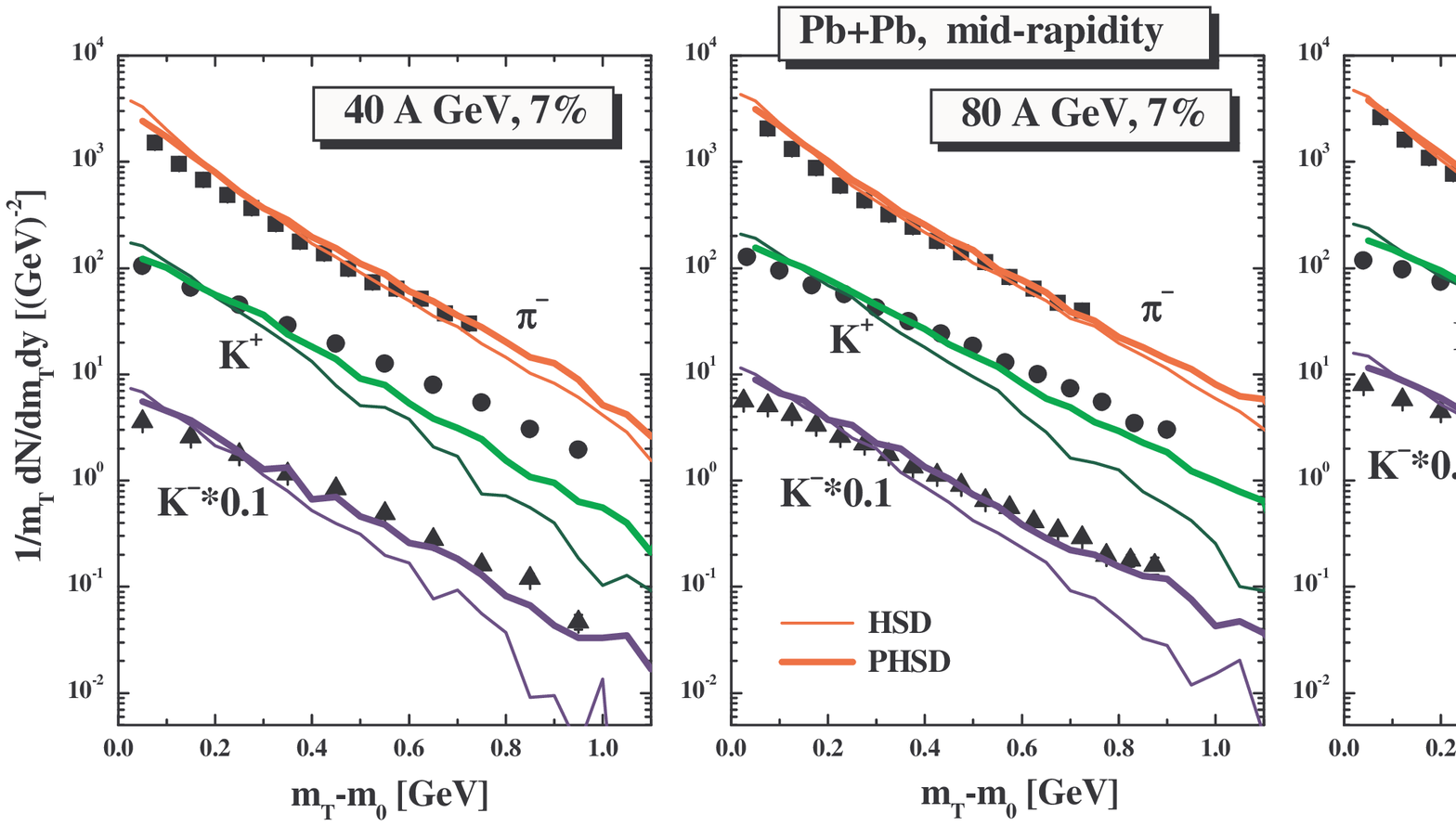,width=0.7\textwidth}} \caption{The
$\pi^-$, $K^+$ and $K^-$ transverse mass spectra for central Pb+Pb
collisions at 40, 80 and 158 A$\cdot$GeV from PHSD (thick solid
lines) in comparison to the distributions from HSD (thin solid
lines) and the experimental data from the NA49 Collaboration
\cite{NA49a}.  } \label{fig14}
\end{figure}

The off-shell parton dynamics also allows for a solution of the
hadronization problem: the hadronization occurs by quark-antiquark
fusion or 3 quark/3 antiquark recombination which is described by
covariant transition rates as introduced in
Refs.~\cite{Cassing:2009vt,CasBrat}, obeying flavor
current-conservation, color neutrality as well as energy-momentum
conservation. Since the dynamical quarks become very massive close
to $T_c$, the formed resonant 'pre-hadronic' color-dipole states
($q\bar{q}$ or $qqq$) are of high invariant mass, too, and
sequentially decay to the ground-state meson and baryon octets
increasing the total entropy. This solves the entropy problem in
hadronization in a natural way~\cite{CassXing}.

\vspace{-20pt}
\section{Particle spectra in comparison to experiment}
\vspace{-10pt} \vspace{-10pt}

\begin{figure}
\centerline{\psfig{figure=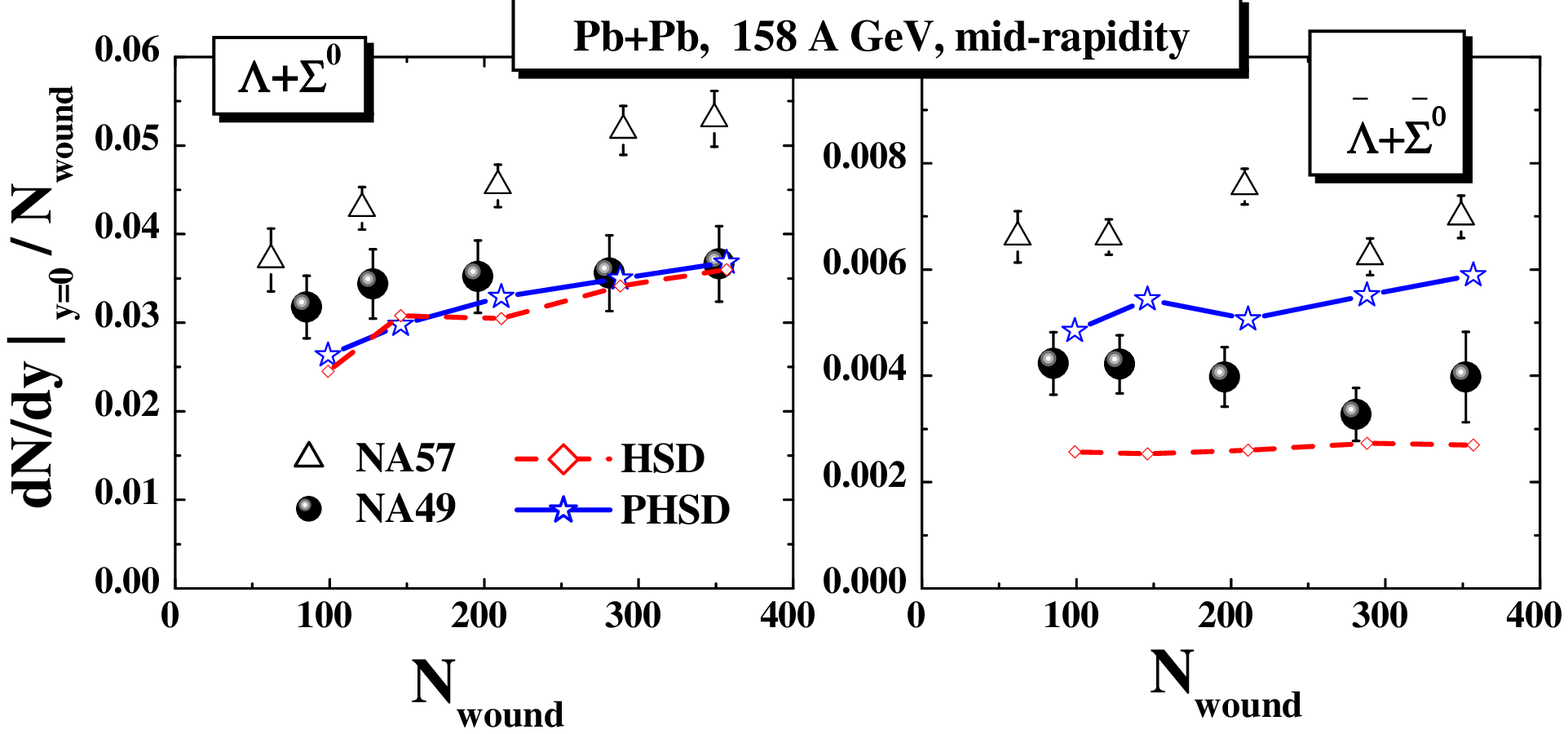,width=0.9\textwidth}}
\centerline{\psfig{figure=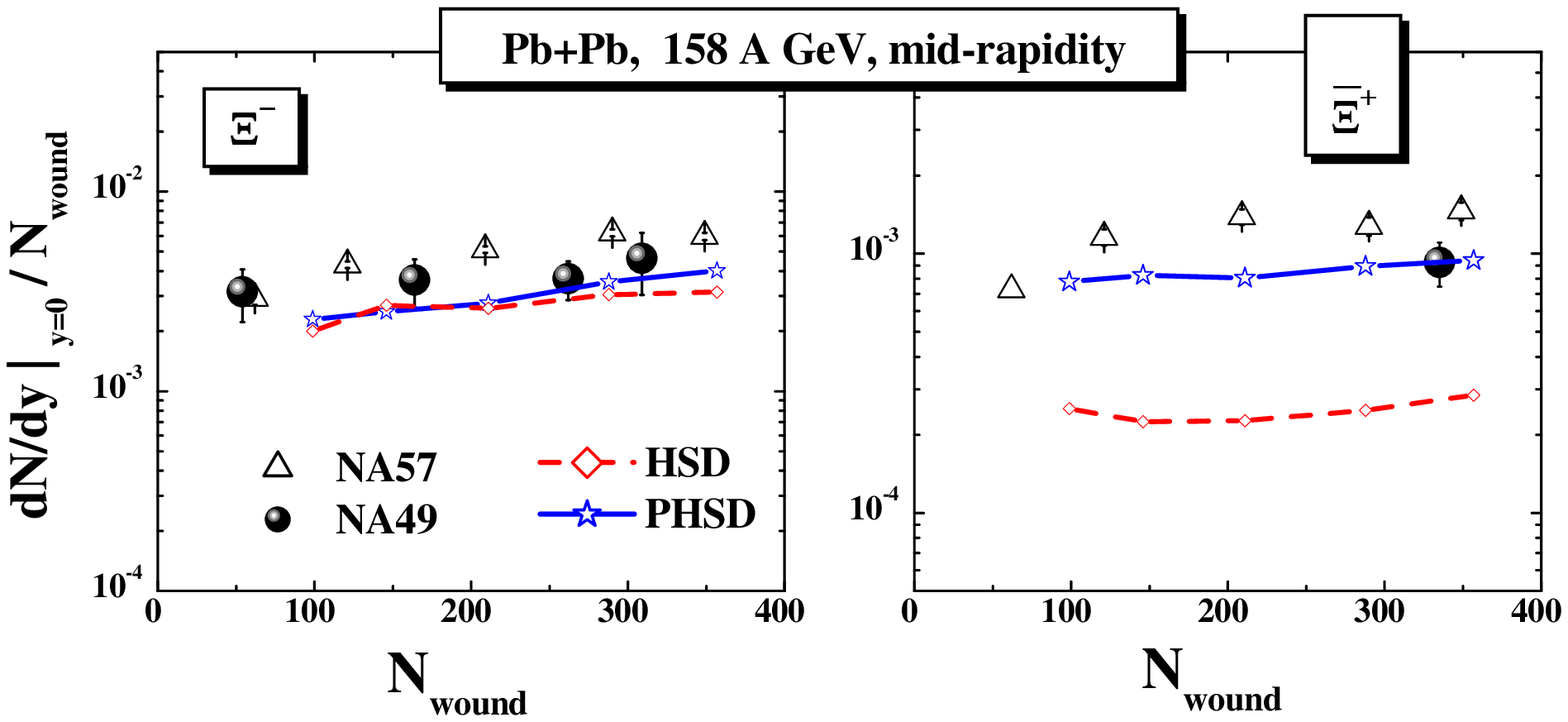,width=0.9\textwidth}}
\caption{{\bf Top:} The multiplicities of $(\Lambda +
\Sigma^0)/N_{wound}$ (l.h.s.) and $(\bar \Lambda + \bar
\Sigma^0)/N_{wound}$ (r.h.s.) as a function of the number of wounded
nucleons for Pb+Pb collisions at 158 A$\cdot$GeV at mid-rapidity
from PHSD (blue solid lines) and HSD (red dashed-dotted lines) in
comparison to the experimental data from the NA57 Collaboration
\cite{NA57} (open triangles) and the NA49 Collaboration
\cite{NA49_aL09} (solid dots). The calculations have an error of
5$-$10\% due to limited statistics. {\bf Bottom:} The multiplicities
of $\Xi^-/N_{wound}$ (l.h.s.) and $\bar \Xi^+/N_{wound}$ (r.h.s.) vs
$N_{wound}$ for Pb+Pb collisions at 158 A$\cdot$GeV at mid-rapidity.
Line coding as above.} \label{fig15d} \label{fig15e}
\end{figure}

It is of interest, how the PHSD approach compares to the
HSD~\cite{HSD} model (without explicit partonic degrees-of-freedom)
as well as to experimental data. In Fig.~\ref{fig14} we show the
transverse mass spectra of $\pi^-$, $K^+$  and $K^-$ mesons for 7\%
central Pb+Pb collisions at 40 and 80 A$\cdot$GeV and 5\% central
collisions at 158 A$\cdot$GeV in comparison to the data of the NA49
Collaboration \cite{NA49a}.  Here the slope of the $\pi^-$ spectra
is only slightly enhanced in PHSD relative to HSD which demonstrates
that the pion transverse motion shows no sizeable sensitivity to the
partonic phase. However, the $K^\pm$ transverse mass spectra are
substantially hardened with respect to the HSD calculations at all
bombarding energies - i.e. PHSD is more in line with the data - and
thus suggest that partonic effects are better visible in the
strangeness-degrees of freedom. The hardening of the kaon spectra
can be traced back to parton-parton scattering as well as a larger
collective acceleration of the partons in the transverse direction
due to the presence of repulsive vector fields for the partons. The
enhancement of the spectral slope for kaons and antikaons in PHSD
due to collective partonic flow shows up much clearer for the kaons
due to their significantly larger mass (relative to pions). We
recall that in Refs. \cite{BratPRL
} the underestimation of the $K^\pm$ slope by HSD (and also UrQMD)
had been suggested to be a signature for missing partonic degrees of
freedom; the present PHSD calculations support this early
suggestion.

The strange antibaryon sector is of further interest since here the
HSD calculations have always underestimated the yield \cite{Geiss}.
In this respect we compare in Fig.~\ref{fig15d} (top part) the
multiplicities of $(\Lambda + \Sigma^0)/N_{wound}$ (l.h.s.) and
$(\bar \Lambda + \bar \Sigma^0)/N_{wound}$ (r.h.s.) as functions of
the number of wounded nucleons $N_{wound}$ for Pb+Pb collisions at
158~A$\cdot$GeV at mid-rapidity from PHSD and HSD to the
experimental data from the NA57 Collaboration \cite{NA57} and the
NA49 Collaboration \cite{NA49_aL09}. Whereas the HSD and PHSD
calculations both give a reasonable description of the $\Lambda +
\Sigma^0$ yield of the NA49 Collaboration, both models underestimate
the NA57 data (open triangles) by about 30\%. An even larger
discrepancy in the data from the NA49 and NA57 Collaborations is
seen for $(\bar \Lambda + \bar \Sigma^0)/N_{wound}$ (r.h.s.); here
the PHSD calculations give results which are in between the NA49
data (solid dots) and the NA57 data (open triangles). We see that
HSD underestimates the $(\bar \Lambda + \bar \Sigma^0)$ midrapidity
yield at all centralities. This observation points towards a
partonic origin but needs further examination.

The latter result suggests that the partonic phase does not show up
explicitly in an enhanced production of strangeness (or in
particular strange mesons and baryons) but leads to a different
redistribution of antistrange quarks between mesons and antibaryons.
To examine this issue in more detail we show in Fig.~\ref{fig15e}
(bottom part) the multiplicities of  $\Xi^-$ baryons (l.h.s.) and
$\bar \Xi^+$ antibaryons (r.h.s.) - devided by $N_{wound}$ - as a
function of the number of wounded nucleons for Pb+Pb collisions at
158 A$\cdot$GeV at mid-rapidity from PHSD and HSD in comparison to
the experimental data from the NA57 Collaboration \cite{NA57} and
the NA49 Collaboration \cite{NA49b,NA49_aL09}. The situation is very
similar to the case of the strange baryons and antibaryons before:
we find no sizeable differences in the double strange baryons from
HSD and PHSD -- in a good agreement with the NA49 data -- but
observe a large enhancement in the double strange antibaryons for
PHSD relative to HSD.

\vspace{-20pt}
\section{Di-jet correlations at RHIC energies} \vspace{-10pt}
\vspace{-10pt}

Di-hadron correlations measure the associated particle distribution
in azimuthal angle $\Delta \phi$ and pseudorapidity $\Delta\eta$
with respect to the high-$p_T$ `trigger' particle.  The data on
two-particle spectra in the high-$p_T$ region in Au+Au collisions
for the c.m.s. energy of the nucleon pair $\sqrt{s}_{NN}=200$~GeV
can be summarized as follows:  1) a strong suppression of the
away-side hadrons (jet quenching)~\cite{Adler:2002tq,phenix}; 2) a
specific `Mach-cone' structure in azimuthal angle $\Delta \phi$ in
the region of the away-side jet ~\cite{phenix}; 3) a long-range in
pseudorapidity $\Delta \eta$ correlation (`ridge') in the region of
the near-side jet \cite{STAR,PHOBOS}. The long-range correlations in
$\Delta \eta$ might be a consequence of string-like correlation
phenomena. In order to explore especially the conjecture of
string-like correlations, we use the HSD model for the study of
di-jet correlations, which employs dominantly early string formation
in elementary reactions and their subsequent decay as well as
hadronic and prehadronic interactions.

\begin{figure}
  \centering
    \begin{minipage}{0.485\textwidth}
      \begin{center}
\epsfig{file=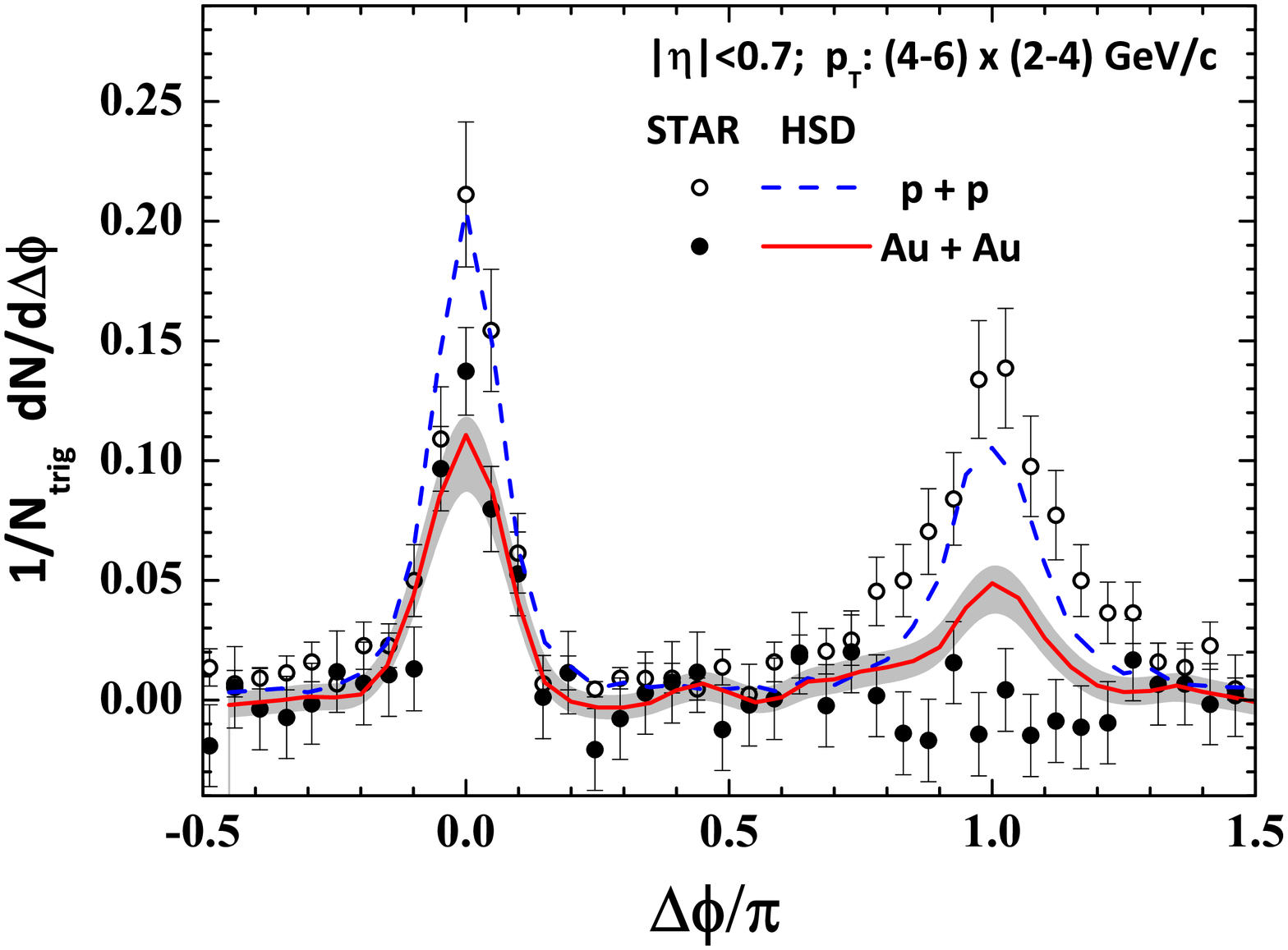,width=\textwidth}
\end{center}
      \caption{Angular correlations of associated particles
($2<p_T^{assoc}<4$~GeV/c) with respect to a trigger particle with
$p_T^{trig}>4$~GeV/c in p+p and in central Au+Au collision events
within the HSD transport approach in comparison to the STAR
data~\cite{Adler:2002tq}. The grey area corresponds to the
statistical uncertainties of the HSD calculations.}
      \label{AngCorr0}
    \end{minipage}
    \hspace{0.01\textwidth}
    \begin{minipage}{0.485\textwidth}
    \begin{center}
\epsfig{file=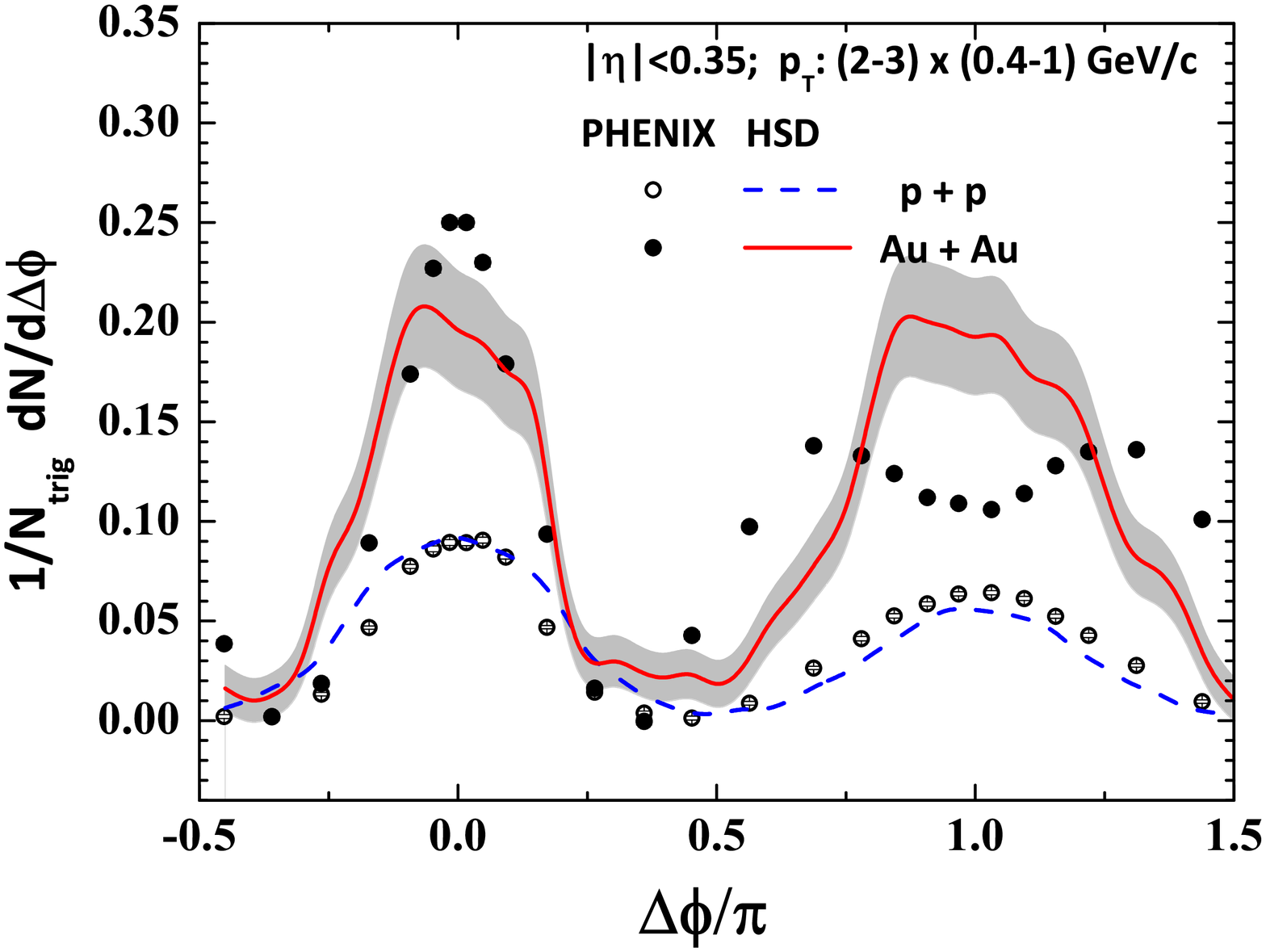,width=\textwidth}
\end{center}
      \caption{Angular correlations of associated particles in p+p and in
central Au+Au collisions within the HSD transport approach in
comparison to the PHENIX data~\cite{phenix} at trigger transverse
momentum $p_T^{trig}=2\div 3$~GeV/c and the associated particle
transverse momentum $p_T^{assoc}=0.4\div 1$~GeV/c. The grey areas
correspond to the statistical uncertainties of the HSD
calculations.}
      \label{AngCorr}
    \end{minipage}
\end{figure}

The previous HSD analysis of high-$p_T$ spectra in
Refs.~\cite{Gallmeister:2004iz,jetHSD} includes all model details
and discusses the nuclear modification factor $R_{AA}(p_T)$ as the
function of $p_T$ and centrality.  In extension of the previous
investigations we now include the full evolution of jets in the
transport approach including the response of the medium, which is
important as the `Mach-cone' and `ridge' structures  are attributed
to medium evolution effects due to jet-medium interactions.  In the
current HSD calculations we use approximately 30$\times 10^6$ of p+p
inelastic collision events and 0.5$\times 10^6$ of central Au+Au
collisions with impact parameter $b=0$. We use the mixed events
method which allows to properly subtract the background by taking
associated particles for each trigger particle from another randomly
chosen event.

\begin{figure}
\epsfig{file=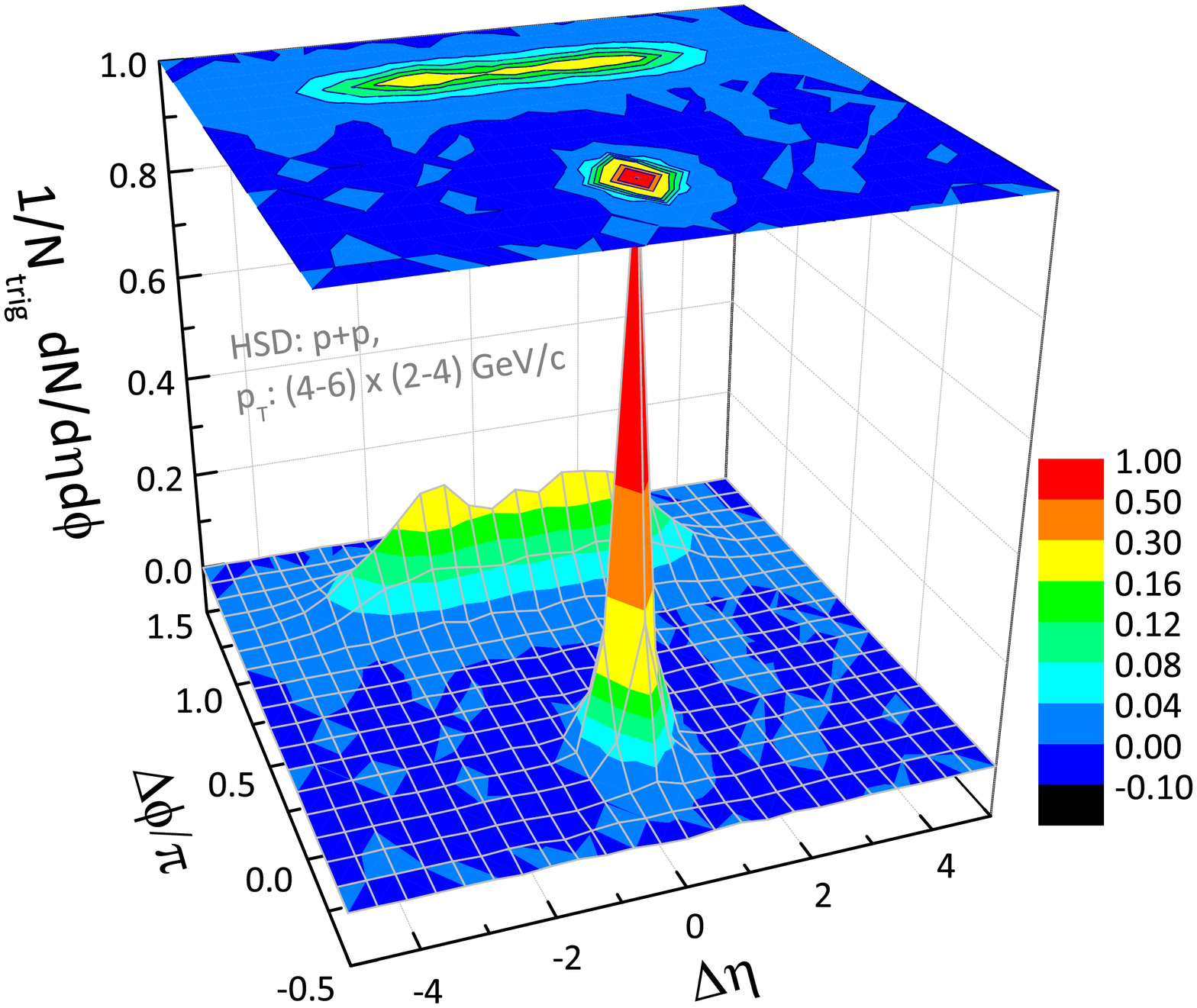,width=0.49\textwidth}
\epsfig{file=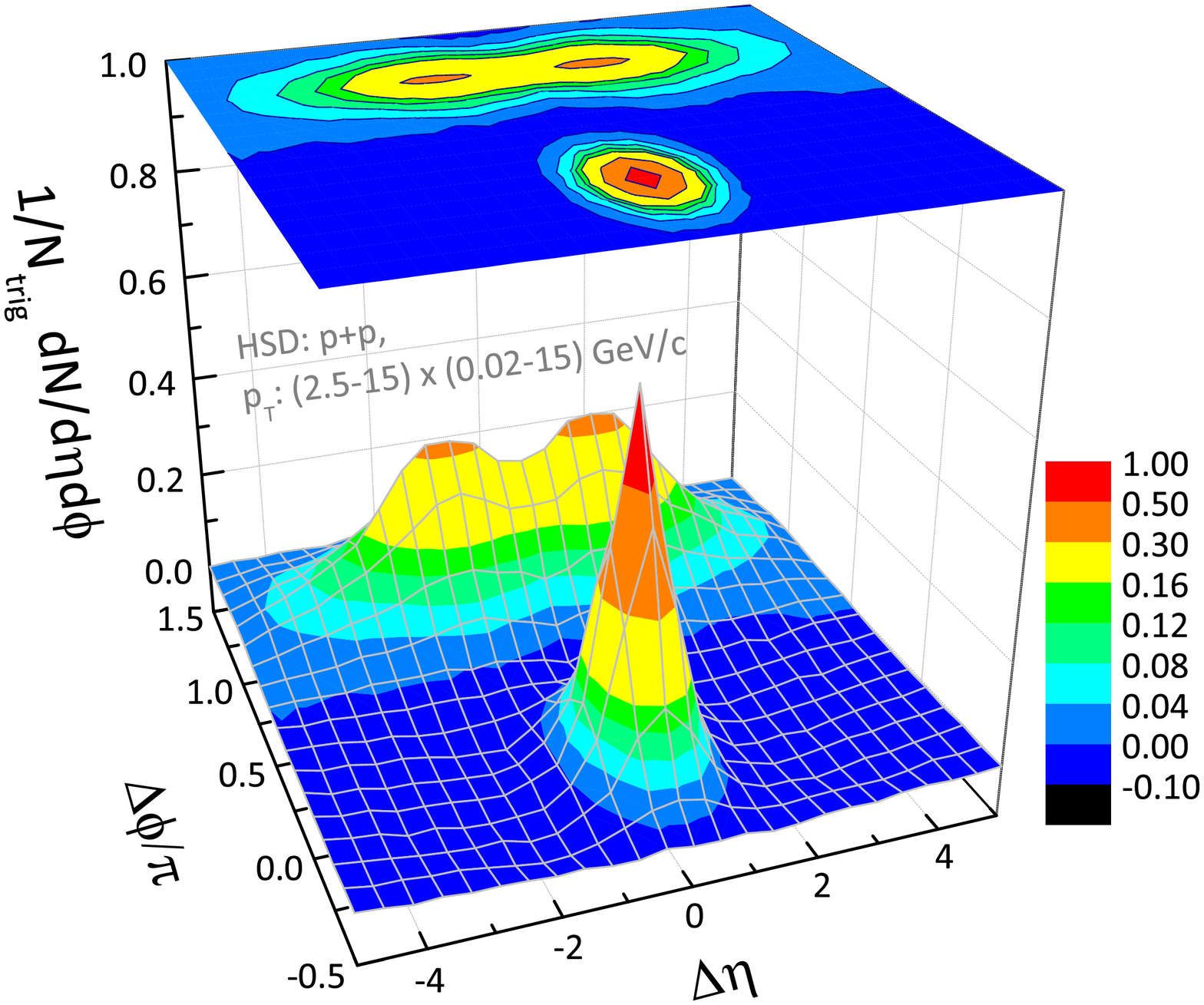,width=0.49\textwidth}
\epsfig{file=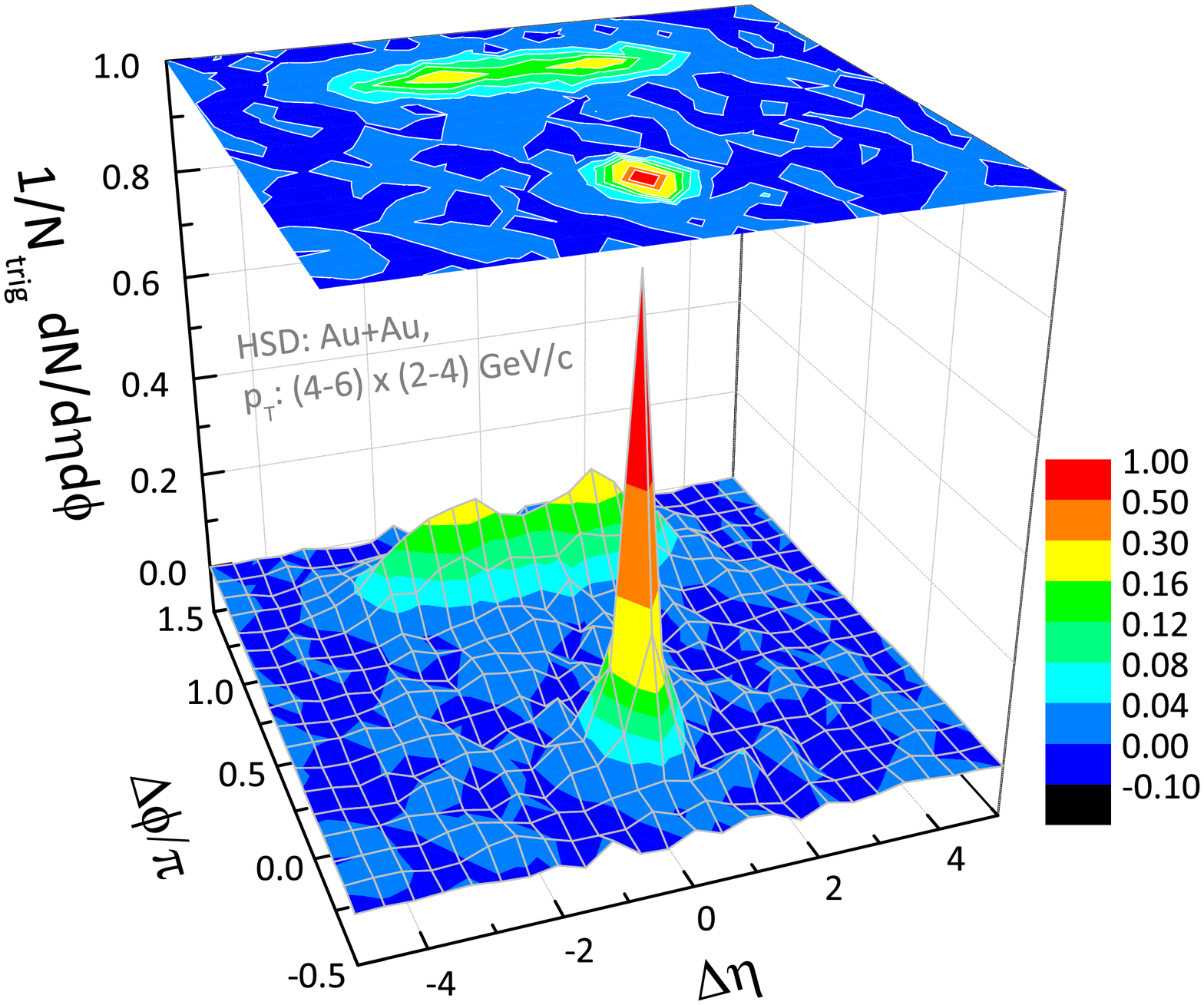,width=0.49\textwidth}
\epsfig{file=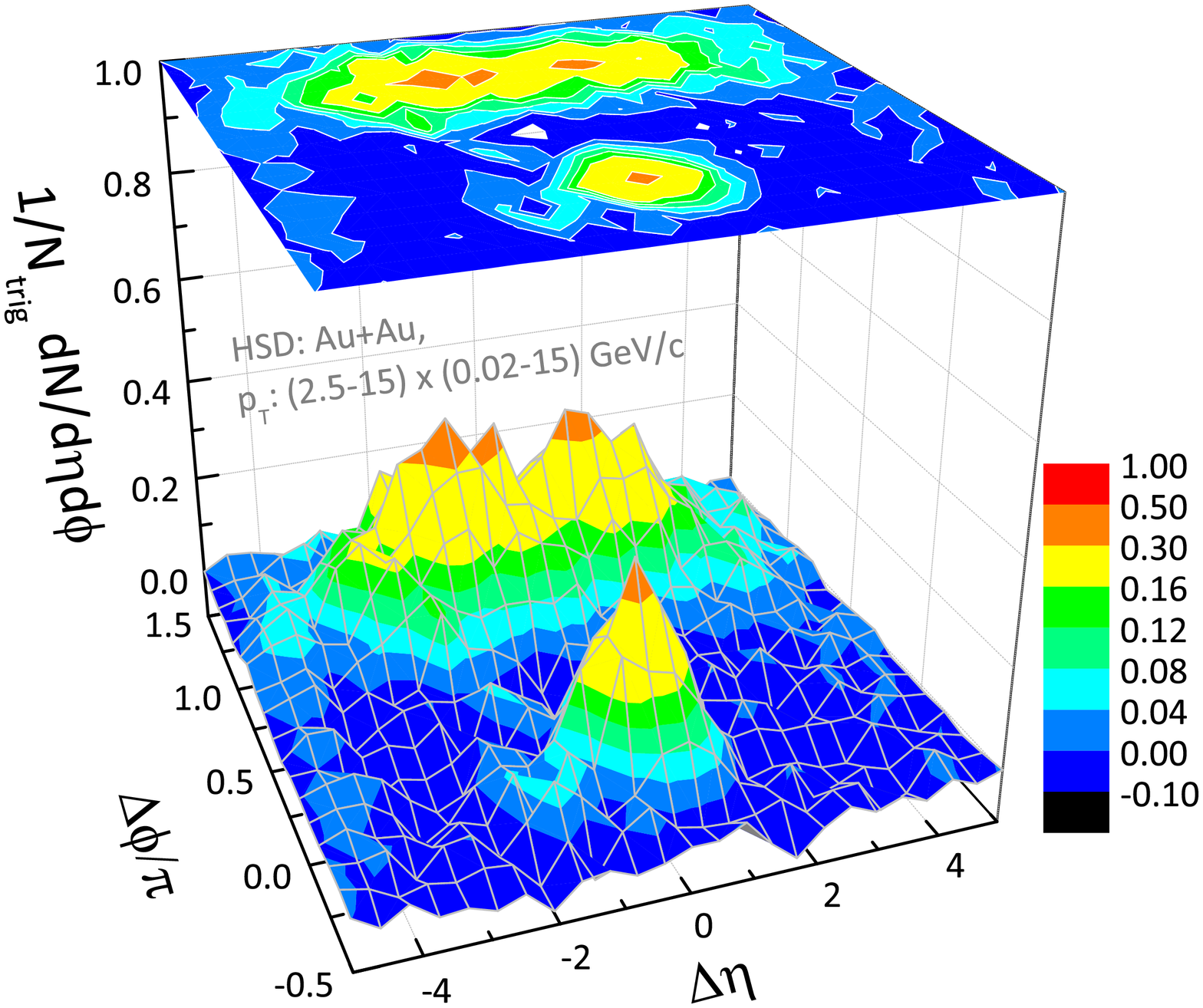,width=0.49\textwidth} \caption{ The
associated particle $(\Delta\eta~,~\Delta\phi)$ distribution
 for p+p ({\it top} panels) and central Au+Au (b=0,
{\it bottom} panels) collisions for the trigger hadron with
$4>p_T^{trig}>6$~GeV/c ({\it left}) and $p_T^{trig}>2.5$~GeV/c ({\it
right}) within the HSD transport approach.} \label{fig3D}
\end{figure}

Angular correlations of associated particles in p+p and central
Au+Au collisions are shown in Figs.~\ref{AngCorr0} and
\ref{AngCorr}. There are two maxima: the `near-side' and `away-side'
peaks at $\Delta\phi=0$ and $\Delta\phi=\pi$, correspondingly.
Fig.~\ref{AngCorr0} provides a comparison to the data from the STAR
Collaboration with the cuts for $p_T^{trig}$ and $p_T^{assoc}$
similar to those in the previous HSD calculations \cite{jetHSD}. We
find a good agreement between the earlier perturbative~\cite{jetHSD}
and current non-perturbative HSD results.  Thus, one may conclude
that a medium modification in this kinematic region of $p_T^{trig}$
and $p_T^{assoc}$ is small in HSD. We mention that the HSD results
reasonably reproduce the data for p+p collisions (within error
bars).  For Au+Au central collisions HSD shows clearly an
insufficient suppression of the `away-side' peak at $\Delta\Phi/\pi
=1$. Note also that for the most central collisions the
experimentally observed suppression of single-particle spectra at
high-$p_T$ can not fully be described by HSD:
$R_{AA}^{HSD}(p_T)=0.35\div 0.4$ whereas $R_{AA}^{exp}(p_T)=0.2\div
0.25$ at $p_T>4$~GeV/c. Our first conclusion is that the
hadron-string medium is too transparant for high-$p_T$ particles (as
already pointed out in \cite{jetHSD}).

Fig.~\ref{AngCorr} corresponds to the data of the PHENIX
Collaboration with different cuts for $p_T^{trig}=2\div 3$~GeV/c and
$p_T^{assoc}=0.4\div 1$~GeV/c.  This is the kinematic region where
one expects a strong medium response to the jet energy loss. The
experimental data show the presence of a `Mach-cone' structure in
azimuthal angle $\Delta \phi$ for the `away-side' jet. This
structure does not appear in the HSD simulations.

%
In Fig.~\ref{fig3D} we present the HSD results for p+p and Au+Au
collisions for the associated differential particle
$(\Delta\eta~,~\Delta\phi)$ distribution.  We use the same cuts as
the STAR Collaboration, $4<p_T^{trig}<6$~GeV/c and
$2<p_T^{assoc}<4$~GeV/c ~\cite{STAR}, and for the PHOBOS
Collaboration, $p_T^{trig}>2.5~GeV$ and
$p_T^{assoc}>0.02$~GeV/c~\cite{PHOBOS}.  In the HSD transport
calculations we obtail on average 0.5 and 5 trigger particles in an
event for the STAR and PHOBOS set of cuts, correspondingly. The
di-jet correlations obtained in the HSD transport simulations of
Au+Au collisions (Fig.~\ref{fig3D}, {\it bottom} panels) do not show
a ridge structure in pseudorapidity $\Delta \eta$ for the near-side
jet as in the data~\cite{STAR,PHOBOS}.

\vspace{-20pt}
\section*{Summary} \vspace{-10pt} \vspace{-10pt}

The PHSD approach has been applied to nucleus-nucleus collisions
from 20 to 160 A GeV in order to explore the space-time regions of
`partonic matter'. We have found that even central collisions at the
top SPS energy of $\sim$160 A GeV show a large fraction of
non-partonic matter. It is also found that though the partonic phase
has only a very low impact on rapidity distributions of
hadrons~\cite{Cassing:2009vt}, it has a sizeable influence on the
transverse-mass distribution of final kaons due to the repulsive
partonic mean-fields and parton interactions. On the other hand, the
most pronounced effect of the partonic phase is seen on the
production of multi-strange antibaryons due to a slightly enhanced
$s{\bar s}$ pair production in the partonic phase from massive
time-like gluon decay and a more abundant formation of strange
antibaryons in the hadronization process. We also mention that
partonic production channels for dileptons appear to be visible in
the $\mu^+ \mu^-$ spectra from In+In collisions at 158~A$\cdot$GeV
in the intermediate invariant mass range~\cite{Linnyk:2009nx}.

Additionally, we conclude that the HSD hadron-string medium does not
show enough suppression for the nuclear modification factor
$R_{AA}(p_T)$ at high $p_T$ and for the away-side jet-associated
particles.  For the first time the medium response on the jet
interactions has been taken into account in the present
non-perturbative HSD calculations in extension to previous
perturbative studies [6]. The non-perturbative calculations,
however, do not reproduce the experimentally observed `Mach-cone'
structure in $\Delta \phi$ for the away-side jet and the long-range
rapidity correlations (the `ridge') for the near-side jet while
supporting the results from the perturbative investigations.  It is
interesting to check in future whether the PHSD model --
incorporating explicit partonic degrees of freedom and dynamical
hadronization   -- will be able to improve an agreement with the
data and reproduce the structures observed by the PHOBOS and STAR
Collaborations.

{\bf Acknowledgments} \vspace{-10pt}

Work supported in part by the ``HIC for FAIR" framework of the
``LOEWE'' program. \vspace{-20pt}
%
%


\begin{thebibliography}{99}
\vspace{-15pt}
\bibitem{CasBrat} W. Cassing and E. L. Bratkovskaya, {Phys. Rev.} C 78, 2008, P. 034919.
\bibitem{Juchem}
W. Cassing and S. Juchem, {Nucl. Phys.} A 665, 2000, P. 377; {\it
ibid} A 672, 2000, P. 417.
\bibitem{Cassing:2009vt}
W.~Cassing and E.~L.~Bratkovskaya,
Nucl.\ Phys.\  A {831}, 2009, P. 215.
\bibitem{Linnyk:2009nx}
  O.~Linnyk, E.~L.~Bratkovskaya and W.~Cassing,
  Nucl.\ Phys.\  A {830}, 2009, P. 491C.
\bibitem{Linnyk:2008hp}
  O.~Linnyk, E.~L.~Bratkovskaya and W.~Cassing,
  Int.\ J.\ Mod.\ Phys.\  E {17}, 2008, P. 1367.
\bibitem{Gallmeister:2004iz}
  K.~Gallmeister and W.~Cassing,
  Nucl.\ Phys.\  A {748}, 2005, P. 241.
\bibitem{KBaym}
      L. P. Kadanoff and G. Baym, {\it Quantum Statistical Mechanics},
      Benjamin, 1962.
\bibitem{Sascha1}
      S. Juchem {\it et al.}, Nucl. Phys. A {743}, 2004, P. 92.
\bibitem{Andre}
      A. Peshier and W. Cassing, { Phys. Rev. Lett.} 94, 2005, P. 172301.
\bibitem{HSD}
      W. Cassing and E. L. Bratkovskaya, { Phys. Rept.} 308, 1999, P. 65;
      W. Ehehalt and W. Cassing, {Nucl. Phys.} A 602, 1996, P. 449.
\bibitem{Cassing06}
      W. Cassing, {Nucl. Phys.} A 791, 2007, P. 365; {\it ibid.}  A 795, 2007,
      P. 70.
\bibitem{CassXing}
 W.~Cassing, E.~L.~Bratkovskaya and Y.~Xing, Prog. Part. Nucl.
 Phys. 62, 2009, P. 359.
\bibitem{NA49a}
C. Alt et al., NA49 Collaboration, Phys. Rev. C 66, 2002, P. 054902;
Phys. Rev. C 77, 2008, P. 024903.
\bibitem{Geiss}  J. Geiss, W. Cassing and C. Greiner, Nucl.
Phys. A 644 , 1998, P. 107.
\bibitem{NA57} F. Antinori {\it et al.}, Phys. Lett. B 595, 2004, P. 68;
  J. Phys. G: Nucl. Phys. 32, 2006, P. 427.
\bibitem{NA49_aL09}
  T.~Anticic {\it et al.}  [NA49 Collaboration],
  Phys.\ Rev.\  C {\bf 80}, 2009, P. 034906.
\bibitem{NA49b}
C. Alt et al., NA49 Collaboration, Phys. Rev. C 78, 2008, P. 034918.
\bibitem{Adler:2002tq}
  C.~Adler {\it et al.}  [STAR Collaboration],
  Phys.\ Rev.\ Lett.\  {90}, 2003, P. 082302.
\bibitem{phenix}
A. Adare et al. [PHENIX collaboration] Phys. Rev. C {78}, 2008, P.
014901.
\bibitem{STAR}
  B. I. Abelev et al., [STAR collaboration] Phys. Rev. C  {80}, 2009, P. 064912;
  M.~van Leeuwen  [STAR collaboration],
  Eur.\ Phys.\ J.\  C {61}, 2009, P. 569.
\bibitem{PHOBOS}
B.~Alver {\it et al.}  [PHOBOS Collaboration],
  Phys.\ Rev.\ Lett.\  {104}, 2010, P. 062301.
\bibitem{BratPRL}
E. L. Bratkovskaya, S. Soff, H. St\"ocker, M. van Leeuwen, and W.
Cassing, Phys. Rev. Lett. 92, 2004, P. 032302;
\bibitem{jetHSD}
  W.~Cassing, K.~Gallmeister and C.~Greiner,
  Nucl.\ Phys.\  A {735}, 2004, P. 277;
  W.~Cassing, K.~Gallmeister and C.~Greiner,
  J.\ Phys.\ G {30}, 2004, P. S801;
\end{thebibliography}
\end{document}